# A metamaterial frequency-selective super-absorber that has absorbing cross section significantly bigger than the geometric cross section

Jack Ng[*], Huanyang Chen[*,†], and C. T. Chan

*Department of Physics, The Hong Kong University of Science and Technology, Clear Water Bay, Hong Kong, China*

**Abstract**

Using the idea of transformation optics, we propose a metamaterial device that serves as a frequency-selective super-absorber, which consists of an absorbing core material coated with a shell of isotropic double negative metamaterial. For a fixed volume, the absorption cross section of the super-absorber can be made arbitrarily large at one frequency. The double negative shell serves to amplify the evanescent tail of the high order incident cylindrical waves, which induces strong scattering and absorption. Our conclusion is supported by both analytical Mie theory and numerical finite element simulation. Interesting applications of such a device are discussed.

## I.     Introduction

It is possible for a particle to absorb more than the light incident on it. Bohren gave explicit examples in which a small particle can absorb better than a perfect black body of the same size. Examples of such type include a silver particle excited at its surface plasmon resonance and a silicon carbide particle at its surface phonon polariton resonance.[1] However, such mechanism is typically limited to very small particles, and for a fixed particle volume, the absorption cross section cannot increase without bound. We shall show in this article that it is possible to design, within the framework of transformation optics,[2,3] a device whose absorption efficiency can be arbitrarily large, at least in principle. We call such a system the super-absorber (SA).

---

[*] These authors contributed equally to this work.
[†] Email: kenyon@ust.hk

Transformation optics[2,3] can be applied to achieve various novel optical illusion effects.[4,5,6,7,8] The SA proposed here is motivated by the design of a super-scatterer, which has scattering cross-section that is bigger than the geometric cross section. The idea of the superscatterer itself can be traced to the pioneering work of Nicorovici *et al.*,[9] who showed using quasi-statics that when the coating shell's dielectric constant of a coated cylinder is negative to that of the core, the cylinder can behave as if the core is enlarged. Pendry and Ramakrishna[10] proposed a perfect cylindrical lens for imaging, through which they introduce the concept of complementary media. Employing the idea of the transformation optics, Leonhardt and Philbin[11] proposed a folded geometry concept for the perfect lens. The folded geometry[11,12] concept was subsequently applied to build the "anti-cloak"[13], the superscatterer,[8,14,15,16] and the cylindrical superlens[17]. In this article, we go one step further to demonstrate that a SA can be built using the folded geometry configuration, with theoretically unlimited absorbing power at one frequency, in the sense that given a fixed particle volume, the absorption cross sections can be made arbitrarily large for the target frequency, at least in principle.

## II. Result and discussion

For simplicity, we shall consider a two-layer cylindrical system illuminated by transverse electric (TE, meaning *E* field along z-axis) waves at normal incidence. The cylindrical system consists of a core and a shell, as depicted in Fig. 1. The core has a radius *a*, and is composed of an isotropic and homogeneous material characterized by constant constitutive parameters $\varepsilon_C$ and $\mu_C$. For generality, we consider the shell (of radius *b*) to be made from anisotropic and graded metamaterials, characterized by $\mu_r(r)$, $\mu_\theta(r)$, and $\varepsilon_z(r)$ (for TE waves, only $\mu_r$, $\mu_\theta$, and $\varepsilon_z$ are relevant).[18] We shall see later that with a proper design, the constitutive parameters for the shell can be made isotropic, and the permeability can even be made homogeneous, i.e. $\mu_r = \mu_\theta = \mu_S$ and $\varepsilon_z(r) = \varepsilon_S(r)$. Such set of constitutive parameters are much easier to fabricate than the anisotropic and inhomogeneous metamaterials that are commonly employed in other proposed devices.

Let us consider a transformation media with a radial mapping $(r,\theta) \to (f(r),\theta)$. The constitutive parameters are given by

$$\frac{\mu_r}{\mu_0} = \frac{\mu_\theta}{\mu_0} = \frac{f(r)}{r}\frac{1}{f'(r)},$$

$$\frac{\varepsilon_z}{\varepsilon_0} = \frac{f(r)}{r} f'(r). \quad (1)$$

For the specific choice of coordinate mapping

$$f(r) = \begin{cases} b^2/r, & a < r < b, \\ r, & r > b, \end{cases} \quad (2)$$

the corresponding constitutive parameters are isotropic and negative:

$$\mu_s = \mu_r = \mu_\theta = -1,$$
$$\varepsilon_s = \varepsilon_z = -b^4/r^4. \quad (3)$$

The Maxwell equation can be solved analytically [8, 13], and the $z$-component of the incident electric field $E_z$ is given by

$$E_z(r,\theta) = \begin{cases} \sum_m \gamma_m^i J_m(k_C r) e^{im\theta}, & r < a \\ \sum_m \left[\alpha_m^i J_m(k_0 f(r)) + \alpha_m^s H_m^{(1)}(k_0 f(r))\right] e^{im\theta}, & a < r < b \\ \sum_m \left[\beta_m^i J_m(k_0 r) + \beta_m^s H_m^{(1)}(k_0 r)\right] e^{im\theta}, & r > b \end{cases} \quad (4)$$

with $k_C = \sqrt{\varepsilon_C}\sqrt{\mu_C}k_0$. After applying the standard boundary conditions, i.e. the continuity of $E_z$ and $H_\theta$ at $r = a$ and $r = b$, the coefficients for the scattered and internal fields are determined to be

$$\beta_m^s/\beta_m^i = -\frac{\sqrt{\varepsilon_e}J_m(k_0 r_e)J_m'(k_e r_e) - \sqrt{\mu_e}J_m'(k_0 r_e)J_m(k_e r_e)}{\sqrt{\varepsilon_e}H_m(k_0 r_e)J_m'(k_e r_e) - \sqrt{\mu_e}H_m'(k_0 r_e)J_m(k_e r_e)},$$

$$\gamma_m^i/\beta_m^i = \frac{\sqrt{\mu_e}\left[H_m(k_0 r_e)J_m'(k_0 r_e) - H_m'(k_0 r_e)J_m(k_0 r_e)\right]}{\sqrt{\varepsilon_e}H_m(k_0 r_e)J_m'(k_c a) - \sqrt{\mu_e}H_m'(k_0 r_e)J_m(k_c a)}, \quad (5)$$

$$\alpha_m^i = \beta_m^i,$$
$$\alpha_m^s = \beta_m^s,$$

where $\varepsilon_e = (a^4/b^4)\varepsilon_C$, $\mu_e = \mu_C$, $k_e = \sqrt{\varepsilon_e}\sqrt{\mu_e}k_0$ and $r_e = b^2/a$. We note that the coefficients for the scattered field ($\beta_m^s$) of the SA are exactly that of a cylinder with permittivity $\varepsilon_e$, permeability $\mu_e$, and radius $r_e$. The shell of the SA can be considered as a magnifying lens that enlarges the core, but this magnifying lens behave in a weird way such that the smaller the object (the core), the bigger the image, since the effective radius $r_e \sim 1/a$.

It can be shown that if the core material for the SA is replaced by a PEC of radius $a$ (which does not absorb), the device will behave as it is a PEC of effective radius of $b^2/a$, resulting in a super-scatterer [8, 16]. Now, with the core material being a lossy dielectric material, we shall see that cylindrical lens amplifies the core to become an object with large absorption and/or scattering cross section, with the details controlled by the core material properties. Here, we are interested in achieve a "super-absorber". Fig. 2 shows the extinction efficiency $Q_{ext}$ and the absorption efficiency $Q_{abs}$ versus $(k_0 b)^2 / k_0 a$ for the SA with $\mu_C = 1 + i$, $\varepsilon_C = (1+i)b^4/a^4$, $k_0 b = 2$ and with $a$ varying. It is evident that the efficiencies increase linearly with $(k_0 b)^2 / k_0 a$, and in principle, one can boost up the efficiencies without bound by increasing $(k_0 b)^2 / k_0 a$. At $k_0 a = 0.05$ and $k_0 b = 2$, $Q_{abs}$=23, which is very high. For comparison, $Q_{abs}$=0.0029 for a homogeneous cylinder with the same $\mu_C$ and $\varepsilon_C$. Clearly, the high absorption of the SA is not due to the high dielectric core, rather the double negative shell plays an important role. Moreover, it is possible for us to fine tune the constitutive parameters of the core to achieve better results or different effects. For example, one may use the parameters $k_0 a = 0.01$, $k_0 b = \sqrt{0.004}$, $\mu_C = 1+i$, $\varepsilon_C = (b^4/a^4)(1+i)$ to build an almost "pure" super-absorber (which does not scatter much), whose $Q_{ext}$=6.6, and $Q_{abs}$=6.08. We note that such a SA in fact mimics a blackbody of six times the size of the original SA, in the sense that it scatters little light ($Q_{sca}$=0.54) but absorbing most light ($Q_{abs}$=6.08) incident on the virtual cylinder of effective radius $k_0 r_e = 0.4 \simeq 6 k_0 b$.

It would be of interest to understand the working mechanism of the SA. We note that Mie theory calculations at $k_0 a = 0.05$ and $k_0 b = 2$ converge only after summing up an unusually large number of angular momentum channels ($|m| \leq 90$) [17]. For even smaller $a$, even more terms are needed and the efficiencies increase rapidly. This is, on one hand, expected, since the SA is effectively a homogeneous cylinder of enlarged radius $b^2/a$=80/$k_0$, which of course requires a large number of terms. However, on the other hand, the unusually large number of terms in the Mie series may appear spurious to the readers, because for such a small scatterer with outer radius $k_0 b = 2$, it usually takes only a few terms in the Mie series to converge. To understand the physics behind this peculiar behavior, one must first understand why in ordinary situation the Mie series converge so rapidly for a small scatterer. Consider an incident high order cylindrical wave with order $m=10$, it is seen from the blue dotted line of Fig. 3 that the field amplitude is sizable for $k_0 r > m$ but decays quickly for $k_0 r < m$. Consequently, in ordinary situation, a small scatterer with size $< m/k_0$ does not overlap with the high order cylindrical wave. This is why a small scatterer cannot scatter the high order waves and only low order terms are needed in the Mie theory. The situation is different if a core material is coated by a negative index shell that has graded refractive index that are impedance matched to the outside. We shall see that it amplifies the evanescent tail of the high order incident wave, restoring it to a sizable amplitude before it hits the absorbing core. The amplification of the evanescent tail of the incident cylindrical wave is demonstrated by the red solid line of Fig. 3. With the SA, the decaying tail of the incident cylindrical wave is amplified again by the shell, and a surface wave is formed at the outer surface of the SA. The incident cylindrical wave, with its amplitude restored by the shell, can now be absorbed by the core. Since double negative materials carry surface waves, they can amplify evanescent waves, which lead to many interesting properties, among which the slab perfect lens (with $\varepsilon = \mu = -1$)[19] is one of the most well-known example. The counterpart in cylindrical geometry requires a device which is complimentary to an empty cylindrical space such that it effectively "annihilates" the annular region of empty space just outside the device. Fortunately, such a device can be designed based on transformation optics, and the coordinate mapping (2) can perform the task.

Since the functionality of transformation media works for one frequency, it is interesting to explore the frequency selectivity of such a device. For that purpose, we need to employ some reasonable frequency dependences of $\varepsilon_s$, $\mu_s$, $\varepsilon_C$, and $\mu_C$. We will take the standard Lorentz model where

$$\varepsilon_S(r) = 1 - \frac{(1-f_S^2)(1+b^4/r^4)}{f^2 - f_S^2 + i\gamma_S f},$$

$$\mu_S = 1 - \frac{2(1-f_S^2)}{f^2 - f_S^2 + i\gamma_S f},$$

$$\varepsilon_C = 1 - \frac{255(f_C^2-1)(1+(256/255)^2)}{f^2 - f_C^2 + i(f_C^2-1)(256/255)f}, \quad (6)$$

$$\mu_C = 1 - \frac{1}{f^2 - 1 + if},$$

with $f_S = 0.5$, $\gamma_S = 10^{-4}$, $f_C = 1.2$, and $f$ is the normalized frequency. The parameters for the SA at the normalized working frequency of $f=1$ are given by $\mu_C = 1+i$, $\varepsilon_C = (1+i)(2/0.5)^4$, $k_0 a = 0.5$, and $k_0 b = 2$. The frequency-dependence of the absorption efficiencies for this assumed dispersion is depicted in Fig. 4. While the exact results will depend on the details of the material dispersion, the salient features in Fig. 4 are rather generic. The absorption cross section peaks sharply at the designed frequency with a non-Lorentzian line shape.

In addition to the analytical Mie theory analysis presented above, our conclusion is also supported by finite element simulation using COMSOL MULTIPHYSICS. Fig. 5(c) shows the field pattern for a homogeneous cylinder illuminated by a plane incident wave ($\lambda = 3.14$ unit, $k_0 = 2$ unit$^{-1}$). The cylinder has a radius of 4 unit ($k_0 c = 8$), and $\varepsilon = \mu = 1+i$. Owing to the matched impedance ($\sqrt{\varepsilon}/\sqrt{\mu} = 1$), the cylinder absorbs a large portion of the incident light. From our previous Mie theory analysis, it is known that the 4 unit-radius-cylinder in Fig. 5(c) can be replaced by a smaller SA. To illustrate the point, we show in Fig. 5(b) the case of a SA with $b = 2$ units, $a = 1$ unit, $\varepsilon_C = (b/a)^4(1+i)$ and $\mu_C = 1+i$. It is clear in Fig. 5(b) and (c) that although the SA is only half the size of the homogeneous cylinder, the field patterns outside the 4 unit-radius-circle are identical in both graphs. To convince the reader that the SA is truly absorbing more light than an ordinary scatterer of the same size, we plotted in Fig. 5(a) the case of a 2 unit-radius-cylinder with $\varepsilon = \mu = 1+i$. Compare Fig. 5(a) and (b), the SA (Fig. 5(b)) absorbs more than an ordinary scatterer of the same size (Fig. 5(a)), as expected. We remark that the white flecks in Fig. 5(b) represent over valued field. They are caused by the surface mode resonances excited at the outer boundary of the SA (see Fig. 3). Finally, we would like to point out that a rough surface is much more absorptive than a cylindrical scatterer of similar size. Consequently by using a rough surface, one will then be able to minimize the scattering and maximize the absorption. Similar idea also works for the SA, and this will be an interesting topic for future research.

We aware that in the literature, there is already an absorptive device known as perfect metamaterial absorber.[7] We stress that the SA proposed here is completely different from the perfect metamaterial absorber. The former is designed to achieve arbitrarily large absorption efficiency, whereas the later is designed to mimic a blackbody and for this reason, it has an absorption efficiency close to one. In fact, it is also possible for us to design a SA such that it behaves as a near ideal blackbody, with a high absorption efficiency and a low scattering efficiency.

A frequency selective "super-absorber" has some plausible applications. For example, we can use a super-absorber of a small geometric size to protect a larger area from unwanted radiation of one particular frequency, leaving other frequencies essentially untouched. It can also have energy efficiency implications. We note that by Kirchhoff's law, a good absorber is also a good radiator. Suppose that we have a thermophotovoltaic (TPV) energy conversion system which consists of a cylindrical heated object surrounded by semiconductor photovoltaic cells. An ordinary heated object will emit in a broad band, but the semiconductor PV cell with a given bandgap can only convert a small part of the energy into electric power since photons with energy less than the bandgap will not be absorbed, and photons with more energy than the bandgap will loss energy as heat. If the heated object is a super-absorber with the operation frequency tuned to the bandgap of the PV cell, the thermal radiation will be greatly enhanced in that particular frequency compared with other frequencies, leading to more efficient energy conversion. The "super-radiator" mechanism can be traced to the fact that the photonic density of states must be greatly enhanced at the designed frequency, and also the negative index lens shell gives a strong coupling of EM wave at the designed frequency from the inside to the outside. This can be compared with the effect of photonic crystals thermal emitters, which can serve to suppress the radiation at the stopband frequencies, while the super-radiator boosts the radiation at a desired frequency. The super-radiator and the photonic crystal can work in tandem to complement each other.

### III. Conclusion

We propose a novel metamaterial device called the SA. We demonstrate using analytical Mie theory and finite element method that the device possesses, in principle, unlimited absorption power in the sense that a given fixed device volume, its absorption cross section can be made arbitrarily large. In the calculations we presented, the SA absorbs over fifty times more than a blackbody of the same cross sectional area (see Fig. 2). More importantly, this "effective blackbody area" can be further enhanced by adjusting the parameters of the super absorber. We also showed that the evanescent wave amplification by double negative material is responsible for the spectacular effects.

A coordinate mapping (2) is introduced, which allows us to build the device using isotropic double negative metamaterial, which are much easier to fabricate experimentally than the anisotropic metamaterial employed in previous studies. We note that other coordinate transformation, such as the one employed in ref. 8, can also be used to build a SA. However, the constitutive parameter will then become anisotropic. One interesting but not fully satisfactory strategy may be worth mentioning. We employed the coordinate mapping of ref. 8, but only this time, we allow the parameter $b_0$ to be a complex number. Using such a coordinate mapping, we successfully achieve unlimited large scattering and absorption cross section, however we pay the price that at least one of the constitutive parameters must contain gain, although the gain can be made arbitrarily small by increasing the parameter $b_0$.

We are free to choose the core's constitutive parameters for the SA. In any case, the SA will behave as an effective homogeneous cylinder of radius $r_e=b^2/a$, with constitutive parameters $\varepsilon_e = (a^4/b^4)\varepsilon_C$, and $\mu_e = \mu_C$. In reality, it could be difficult for the experimentalists to fabricate a metamaterial with the desired constitutive parameters. Here, the proposed SA here can map a dielectric constant $\varepsilon_C$ to $\varepsilon_e$. Consequently, in circumstances where one cannot fabricate a metamaterial with $\varepsilon_e$, one may instead fabricate a metamaterial of $\varepsilon_C$, and then map it to $\varepsilon_e$ via our SA recipe. We stress that our recipe works not just for the dielectric constant, for example, if one considers a transverse magnetic wave, it is then the permeability that is being scaled or mapped.

The mechanism behind the spectacular effects is the evanescent wave amplification by double negative material. High order incident cylindrical waves are not scattered or absorbed by an ordinary small scatterer as they have little overlap. For this reason its scattering and absorption cross section is limited. However for the super-absorber, the double negative shell can amplify the evanescent tail of the high order incident cylindrical waves, restoring the waves to a sizable amplitude before they hit the absorptive core, which induces strong scattering and absorption.

Last but not least, in addition to an academic exercise where we demonstrated theoretically that the possibility of unlimited absorption power, we hope that the device can be useful as a frequency-selective absorber or as a "super-radiator" that may have energy efficiency implications.

**Acknowledgment**

This work was supported by Hong Kong central allocation Grant No. HKUST3/06C. Computation resources are supported by the Shun Hing Education and Charity Fund.

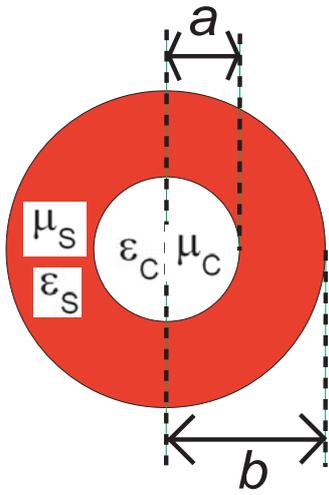

**Fig. 1** The geometry of the SA, which consists of a core of radius *a* and a shell of radius *b*. The core is characterized by constant constitutive parameters $\varepsilon_C$ and $\mu_C$, and the shell is characterized by $\mu_s = -1$ and $\varepsilon_s = -b^4/r^4$.

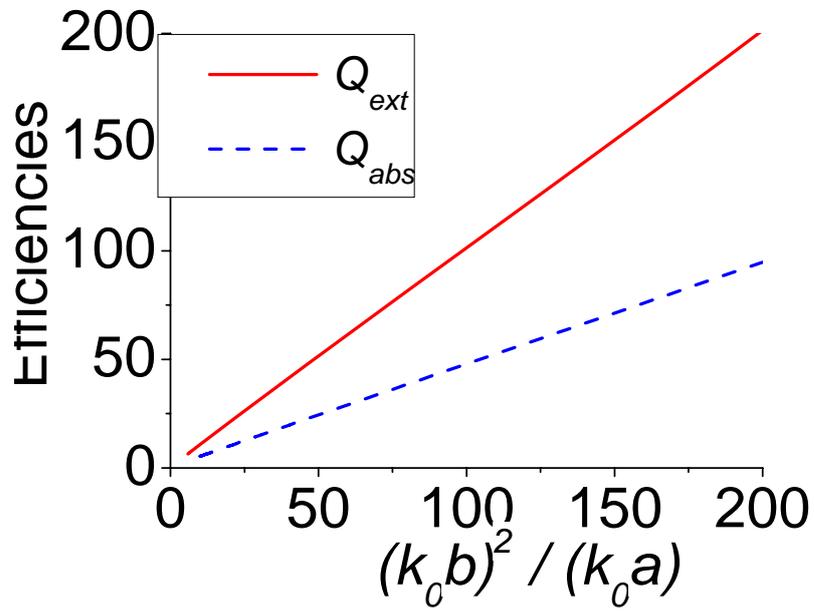

**Fig. 2** The extinction and absorption efficiencies for the SA, with core constitutive parameters $\mu_c = 1+i$, $\varepsilon_c = (1+i)b^4/a^4$ and $k_0 b = 2$.

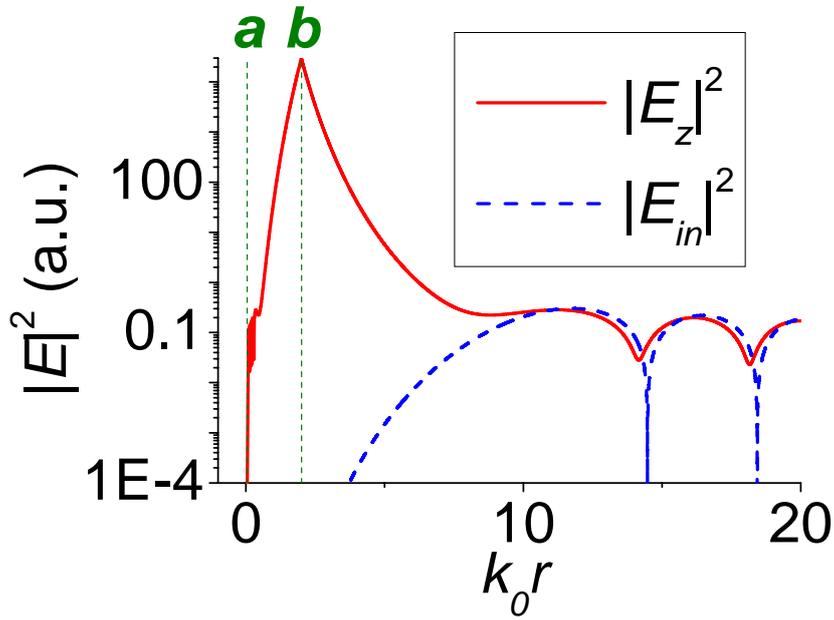

**Fig. 3** The incident field intensity $|E_{in}|^2$ and total field intensity $|E_z|^2$ for an incident cylindrical wave with azimuthal number *m*=10, and the SA has $k_0 a = 0.05$ and $k_0 b = 2$. The intensities are independent of $\theta$, since only one *m* is involved. The inner and outer radii of the SA are indicated by a pair of green lines. Evidently, the evanescent tail of the incident cylindrical wave at small *r* is being amplified by the SA, resulting in strong interaction between the core and the incident wave.

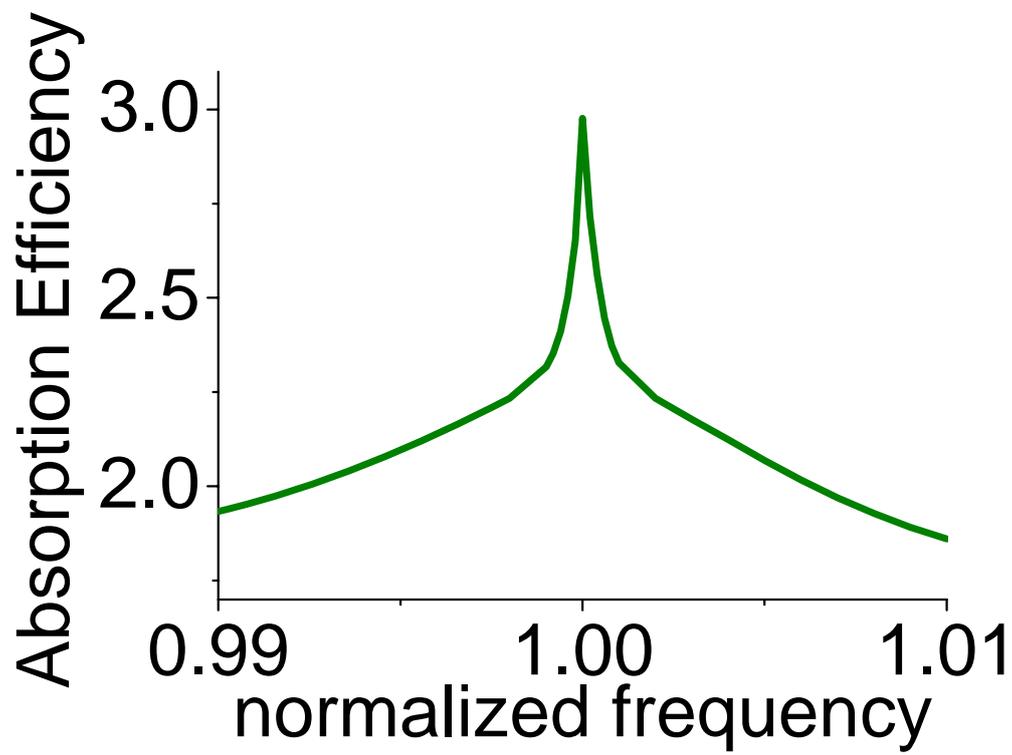

**Fig. 4** The frequency-dependence of the absorption efficiency.

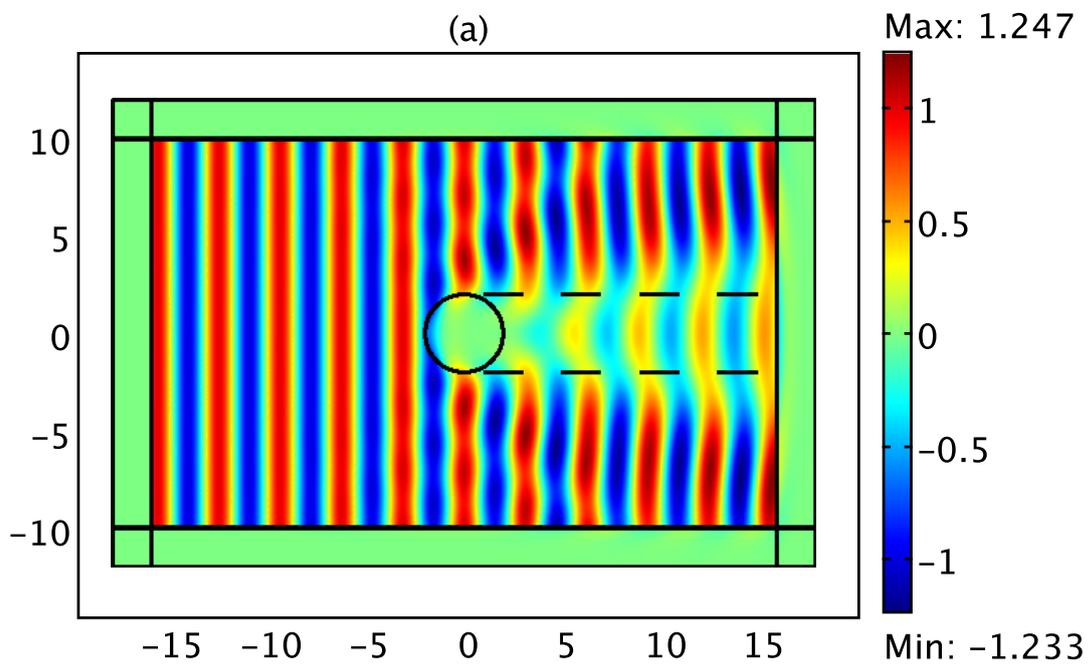

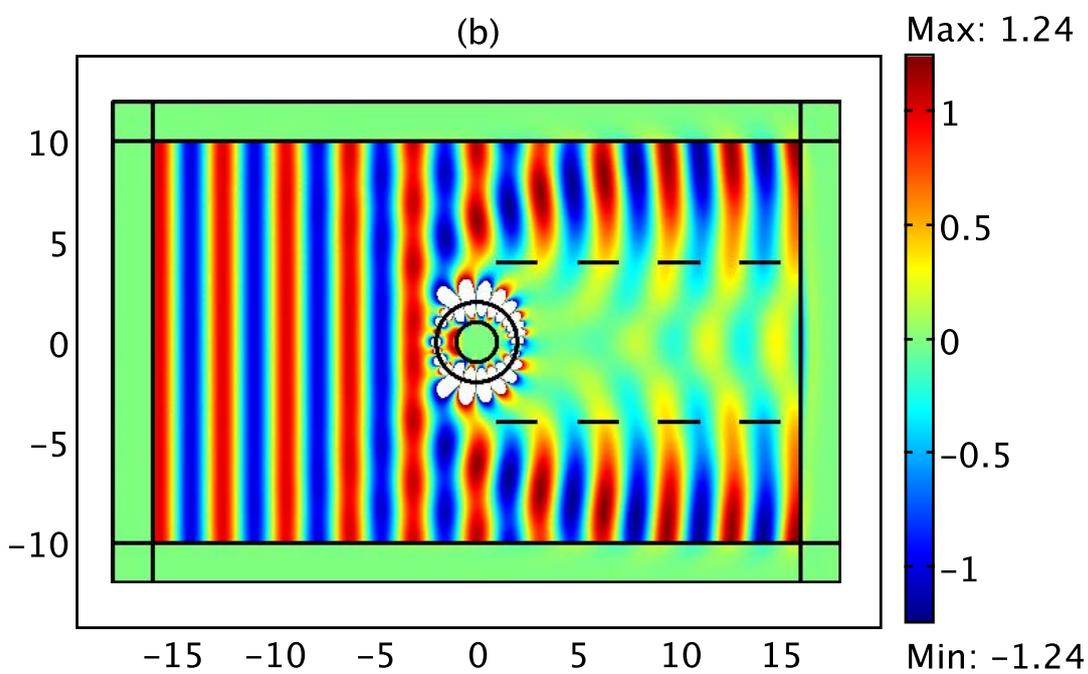

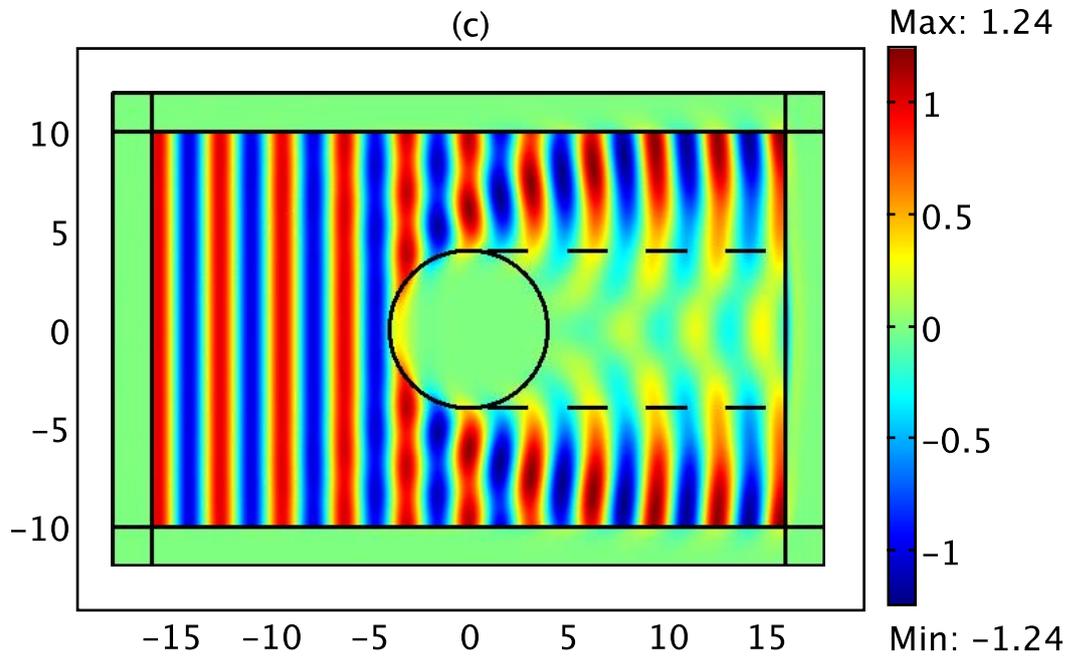

**Fig. 5** (a) The scattering pattern of a homogeneous cylinder of radius $b$=2 unit with $\varepsilon = \mu = 1+i$. (b) The scattering pattern of the SA with inner radius $a$=1unit and outer radius b=2 unit. (c) The scattering pattern of a homogeneous cylinder with radius $b^2/a$=4 unit.

[1] C. F. Bohren, *Am. J. Phys.* **51**, 323 (1983).

[2] J. B. Pendry, D. Schurig, and D. R. Smith, *Science* **312**, 1780 (2006).

[3] U. Leonhardt, *Science* **312**, 1777 (2006).

[4] D. Schurig, J. J. Mock, B.J. Justice, S. A. Cummer, J. B. Pendry, A. F. Starr, and D. R. Smith, *Science* **314**, 977 (2006).

[5] H. Y. Chen and C. T. Chan, *Appl. Phys. Lett.* **90**, 241105 (2007).

[6] M. Rahm, D. Schurig, D. A. Roberts, S. A. Cummer, D. R. Smith, and J. B. Pendry, *Photon. Nanostruct.: Fundam. Applic.* **6**, 87 (2008).

[7] N. I. Landy, S. Sajuyigbe, J. J. Mock, D. R. Smith, and W. J. Padilla, *Phys. Rev. Lett.* **100**, 207402 (2008).

[8] T. Yang, H. Y. Chen, X. Luo, and H. Ma, *Opt. Express* **16**, 18545 (2008).

[9] N. A. Nicorovici, R. C. McPhedran, and G. W. Milton, *Phys. Rev. B* **49**, 8479 (1994).

[10] J. B. Pendry and S. A. Ramakrishna, *J. Phys.: Condens. Matter* **15**, 6345 (2003).

[11] U. Leonhardt and T. G. Philbin, *New J. Phys.* **8**, 247 (2006).

[12] G.W. Milton, N. P. Nicorovici, R. C. McPhedran, K. Cherednichenko, and Z. Jacob, to appear in *New J. Phys.* (2008), see also in the preprint: http://arxiv.org/abs/0804.3903.

[13] H. Y. Chen, X. Luo, H. Ma, and C. T. Chan, *Opt. Express* **16**, 14603 (2008).

[14] H. Y. Chen, X. H. Zhang, X. Luo, H. Ma, and C. T. Chan, to appear in *New J. Phys.* (2008), see also in the preprint: http://arxiv.org/abs/0808.0536.

[15] X. Luo, T. Yang, Y. Gu, and H. Ma, http://arxiv.org/abs/0809.1823.

[16] Y. Luo, J. Zhang, H. S. Chen, B.-I. Wu, and J. A. Kong, http://arxiv.org/abs/0808.0215.

[17] M. Yan, W. Yan, and M. Qiu, *Phys. Rev. B* **78**, 125113 (2008).

[18] Note that the permittivity tensor and permeability tensor are respectively $\vec{\vec{\varepsilon}} = \varepsilon_r \hat{r}\hat{r} + \varepsilon_\theta \hat{\theta}\hat{\theta} + \varepsilon_z \hat{z}\hat{z}$ and $\vec{\vec{\mu}} = \mu_r \hat{r}\hat{r} + \mu_\theta \hat{\theta}\hat{\theta} + \mu_z \hat{z}\hat{z}$.

[19] J. B. Pendry, *Phys. Rev. Lett.* **85**, 3966 (2000).